\documentclass[final,3p,times,twocolumn]{elsarticle}

\usepackage{graphicx,bm,epsf,float,amssymb}

\journal{Nuclear Instruments and Methods}

\begin{document}

\begin{frontmatter}



\title{Directional Fast Neutron Detection Using a Time Projection Chamber}

\author[llnl]{N.~S.~Bowden\corref{cor1}}
\cortext[cor1]{Corresponding Author. Tel.: +1 925 422 4923.}
\ead{nbowden@llnl.gov}

\author[llnl]{M. Heffner}
\author[llnl]{G. Carosi}
\author[llnl]{D. Carter}
\author[rut]{P. O'Malley}
\author[purdue]{J. Mintz}
\author[purdue]{M. Foxe}
\author[purdue]{I. Jovanovic}
\address[llnl]{Lawrence Livermore National Laboratory, Livermore, CA~94550, USA}
\address[rut]{Department of Physics and Astronomy, Rutgers, The State University of New Jersey, Piscataway, NJ 08854, USA}  
\address[purdue]{School of Nuclear Engineering, Purdue University, West Lafayette, IN 47907, USA}

\begin{abstract}
Measurement of the three dimensional trajectory and specific ionization of recoil protons using a hydrogen gas time projection chamber provides directional information about incident fast neutrons. Here we demonstrate directional fast neutron detection using such a device. The wide field of view and excellent gamma rejection that are obtained suggest that this device is well suited to searches for special nuclear materials, among other applications.


\end{abstract}


\end{frontmatter}

\section{Introduction}
\label{sec:intro}

Fast neutron imaging is a promising technique for Special Nuclear Material (SNM) search and safeguards applications due to the penetrating nature of fast neutrons and since naturally occurring background rates are low~\cite{Gordon}. Several techniques have been proposed for the directional detection of fast neutrons for these applications. These include Time Projection Chambers (TPCs) filled with H$_2$~\cite{nTPC1,nTPC2}, $^3$He gas~\cite{Goddard}, or a mixture of gases~\cite{MIT}. Considerable progress has also been made in adapting organic scintillator multiple proton recoil imagers designed for solar neutron imaging~\cite{Ryan} to the lower energies typical of fission neutron sources~\cite{BNL,SNL,Bravar}.

Here we describe the design, operation, and performance of a H$_2$ filled TPC optimized for fast neutron detection. As in all neutron detectors~\cite{Knoll}, we observe the product of a nuclear reaction between an incident neutron and a nucleus in the target medium -- in this case protons that have undergone elastic scattering. However, not only can we detect the occurrence and energy of such a scatter, we can also track in three dimensions (3D) the direction and specific ionization of the recoiling proton. This novel capability provides significant information about the incident fast neutron direction, especially when averaged over several independent scattering events.  

We demonstrate several features of this device that are of particular relevance to SNM search and safeguards applications. First, the ability to localize sources at tens of meters standoff provides a clear benefit over non-imaging detectors, as well as providing an improvement in signal to background. Second, the TPC exhibits a wide field of view -- in principle $4\pi$, and in practice very nearly so -- which is a feature of obvious benefit in a search application. Finally, the TPC possesses very good particle discrimination ability, particularly against gamma rays.

\section{Theory of Operation}

The TPC~\cite{TPC,TPC2} was invented in the 1970's and has been used extensively in fields as diverse as accelerator physics and dark matter searches.  The TPC is capable of producing a full 3D reconstruction of the momentum and ionization produced by a charged particle in its active volume.  Here we describe the application and optimization of this powerful detection technique to the task of fast neutron imaging.

The charged particle tracking provided by TPCs can be used for directional neutron detection via several techniques. Tracking the charged reaction products of a capture reaction (e.g. $^3$He(n,p)$^3$H) yields the incident neutron direction via the vector sum of the reaction product momenta -- knowledge of the reaction $Q$-value also yields the incident energy. Similarly, recording the momentum vectors of two recoiling particles set in motion via elastic scatters by the same neutron enables a complete kinematic reconstruction of that neutron's incident direction and energy. Measurement of only one track and the vertex of the second recoil in a double elastic scatter constrains the incident neutron direction to lie upon a line. Finally, measurement of the momentum vector of a single recoiling particle from an elastic scatter provides a less complete constraint upon the incident neutron direction. In this situation, the tightest constraint is provided when the recoiling particle is a proton, because its mass is similar to that of a neutron: the incident neutron direction is constrained to lie in the hemisphere behind the recoil direction. Furthermore, it is worth noting that averaging an ensemble of single elastic proton scatters can fairly quickly determine the direction of a neutron source~\cite{nTPC2}. Given the relatively high efficiency for a single scatter occurring, this approach appears attractive for rapid SNM discovery and search applications. This is the measurement mode that we have optimized for in this work.

Among hydrogen bearing gases, alkanes possess the advantage of high hydrogen density as a function of gas pressure. This is advantageous in weight sensitive applications, since operation at a lower pressure allows a lighter pressure vessel, or indeed only a non-permable gas envelope at atmospheric pressure. Alkanes are also commonly transported in pressurized form in commerce. 

H$_2$ has previously been used in only a handful of TPCs \cite{hTPC1,hTPC2,hTPC3}. Use of pure H$_2$ as the TPC gas for this application has the advantage that no carbon atoms are introduced into the TPC active volume. These increase the electron density, and therefore reduce track length making precise directional measurements more difficult, and provide large mass scattering centers that can result in large neutron deflections. For a fixed number density of hydrogen atoms, H$_2$ will result in the longest track and therefore provide the lowest energy threshold and most precise trajectory measurement. We have therefore chosen to use a gas primarily composed of H$_2$ for this proof of principle demonstration. To ensure good gas gain stability we use a gas mixture composed of $90\%$ H$_2$ and $10\%$ CH$_4$ (by volume).

Of course, there are several other gases that could be used, but each has significant drawbacks. $^4$He is readily available and non-flammable, but the average recoiling nucleus carries less energy and is not as well aligned with the neutron direction as is the case with a recoiling proton. Furthermore, the larger specific ionization of the recoiling $^4$He nucleus would result in relatively short tracks. Finally, any radon or other alpha-decaying contaminant in the counting gas or vessel materials could produce a high background indistinguishable from neutron induced recoils. $^{10}$BF$_3$ could be used in place of scarce $^3$He, but it is toxic, and again the neutron capture reaction products are heavy, resulting in short, difficult to measure particle tracks.

\section{Detector Design}
\label{sec:design}

\begin{figure}[tb]
\centering
\includegraphics*[width=3in]{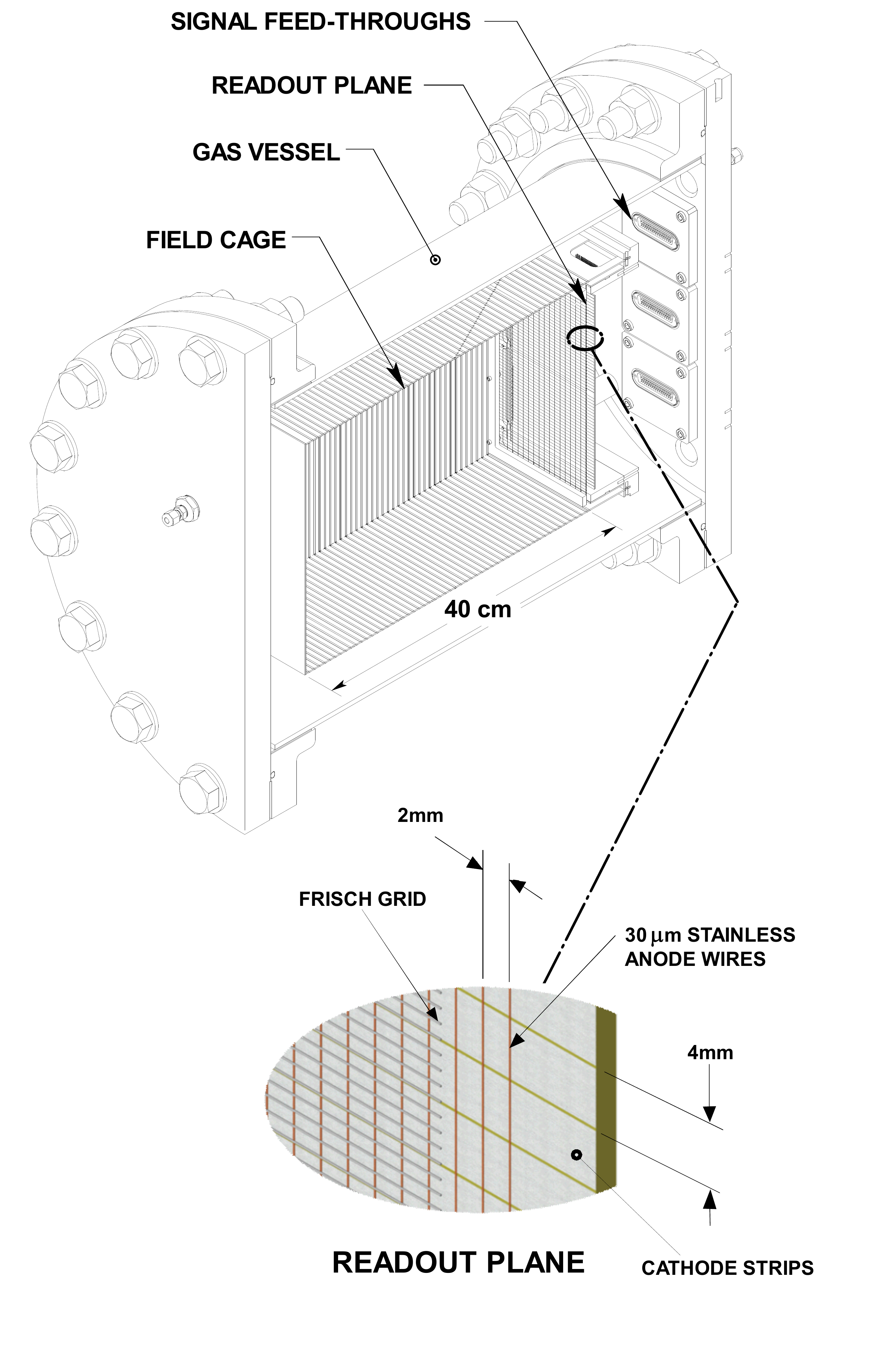}
\caption{A schematic diagram of the TPC.} \label{fig:detector}
\end{figure}

A schematic diagram of the TPC system is shown in Fig.~\ref{fig:detector}. A stainless steel gas vessel ($50$~cm length, $46$~cm diameter) encloses a field cage and charge readout plane. To allow flexible operation of this prototype system, the gas vessel was designed for operation at pressures up to $10$~atmospheres. To support this maximum pressure the vessel wall is $6$~mm thick, with end flanges of $3.8$~cm thickness. As only non-mobile laboratory use was envisioned, no effort was expended to minimize the vessel mass.

The field cage used to establish the drift field comprises five FR-4 Printed Circuit Board (PCB) sheets with conductive strips etched upon them. One end is maintained at a constant potential ($-17.3$~kV), while a resistive chain along the field cage establishes the drift field of $315$~V/cm. The field cage defines an active gas volume of  $25~$cm $\times~25~$cm $\times~40~$cm $=25$~liters. 

Ionization charge is drifted to an amplification and readout region that begins with a Frisch grid, maintained at $-3.6$~kV. This is followed by a linear array of $128$~copper coated stainless steel wires of $30~\mu$m diameter, orientated perpendicular to the horizontal plane of the TPC. A detailed study found that the TPC working gas would not embrittle these fine wires~\cite{nTPC1}. The wires are spaced at $2$~mm intervals, and are held at ground potential. Behind the wire plane is PCB with $64$~etched strips, each $4$~mm apart, orientated parallel to the TPC horizontal plane. The strip plane is also maintained at $-3.6$~kV. 

The crossed wire and strip planes constitute the two dimensional charge readout of the TPC. Both wires and strips are connected to charge sensitive preamplifiers, directly in the case of the wires, and via de-coupling capacitors for the strips. Due to a problem with the particular connection scheme adopted during the data taking period reported here, only every other the strip could be read out, resulting in an effective $8$~mm pitch.

The TPC pressure vessel contains fitting for gas inlet and outlet, as well as pressure and temperature measurement. The system is operated in a constant pressure mode using a PID controller to regulate the position of a needle valve at the gas inlet. A port opposite the readout plane allows for the introduction of a $^{242}$Cm alpha particle source that is used for calibration studies.

For the work described here, the TPC was operated at a pressure of $2$~atmospheres, using a $90\%$ H$_2$/$10\%$ CH$_4$ gas mixture. This operating pressure resulted in track lengths for $0.5$~MeV recoil protons of about $20$~mm -- well matched to the readout pitch and triggering scheme, and providing reasonable detection efficiency for an incident fission neutron spectrum ($\approx 40\%$ of recoils generated by an incident fission neutron spectrum have an energy greater than $0.5$~MeV).

\section{Data Acquisition}
\label{sec:daq}

Signals produced in the charge-sensitive preamplifiers are digitized using a VME data acquisition system (DAQ) that is controlled using the ORCA data acquisition software \cite{orca}. The data acquisition hardware consists of nine 22-channel Waveform Digitizer/First Level Trigger (WFD/FLT) cards and a single Second Level Trigger/Interface card. These were originally produced by FZ Karlsruhe for the fluorescence telescopes of the AUGER experiment \cite{auger}. The WFD cards sample at $10$~MHz and record $1000$~$12$~bit samples per event, i.e. record a total of $100~\mu$s per event.

Each channel has an independent discriminator, which is applied to the average of $5$~samples. The waveforms of channels above threshold are recorded if $5$ or more channels exceed threshold within $4~\mu$s. The discriminator averaging, discriminator thresholds, trigger multiplicity and coincidence time can all be adjusted. The trigger parameter selections used here were found to be suitable for reducing spurious triggers on electronic noise, while retaining a high efficiency for neutron interactions in the TPC. However, as a consequence of the coincidence  requirement, short tracks at certain angles are not able to trigger data acquisition. This results in a reduction in field of view of $10\%$ (Sec.~\ref{sec:quality}). We note that this is a consequence of the particular triggering scheme adopted for this proof of principle demonstration -- a triggering and charge readout scheme sensitive to single channels, and the charge arrival time on that one channel would yield full $4\pi$ sensitivity.

\section{Event Reconstruction}
\label{sec:analysis}

\begin{figure}[!tb]
\centering
\includegraphics*[width=3in]{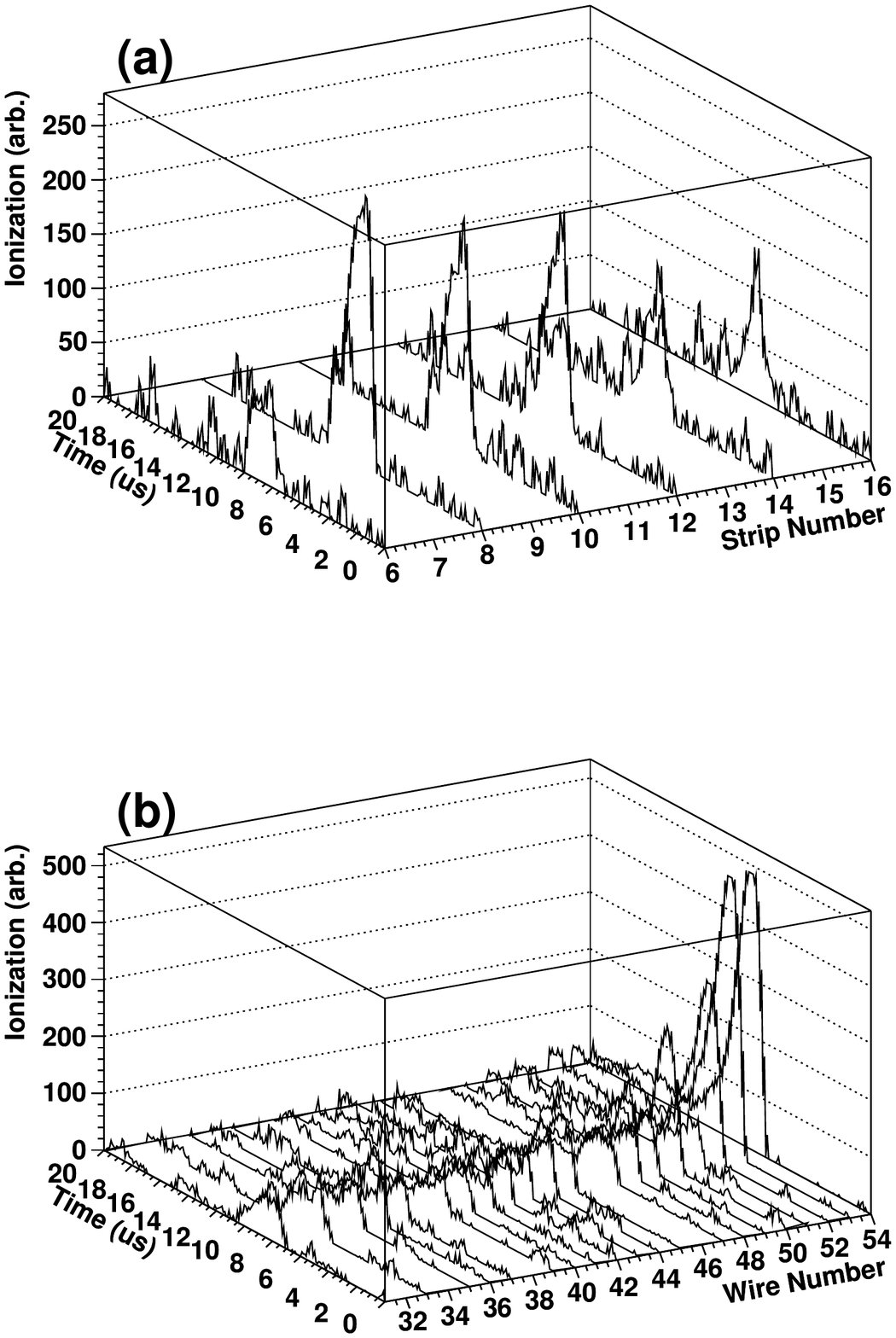}
\caption{The background subtracted waveforms of a typical event for (a) strips and (b) wires.} \label{fig:waveforms}
\end{figure}

\subsection{Waveform Conditioning}

Waveforms are first tested for ADC saturation, and the event rejected if it is found. The overall gain of the system is carefully adjusted so that neutron tracks do not cause such saturation - small discharges or $\alpha$-particle tracks at steep angles with respect to the readout plane are rejected by this selection. Next, a baseline subtraction is performed, using the average of the first and last $100$~samples of each trace. An example of the waveforms that form a typical event are shown in Fig.~\ref{fig:waveforms}.

\subsection{Hit Finding}

\begin{figure}[!tb]
\centering
\includegraphics*[width=3in]{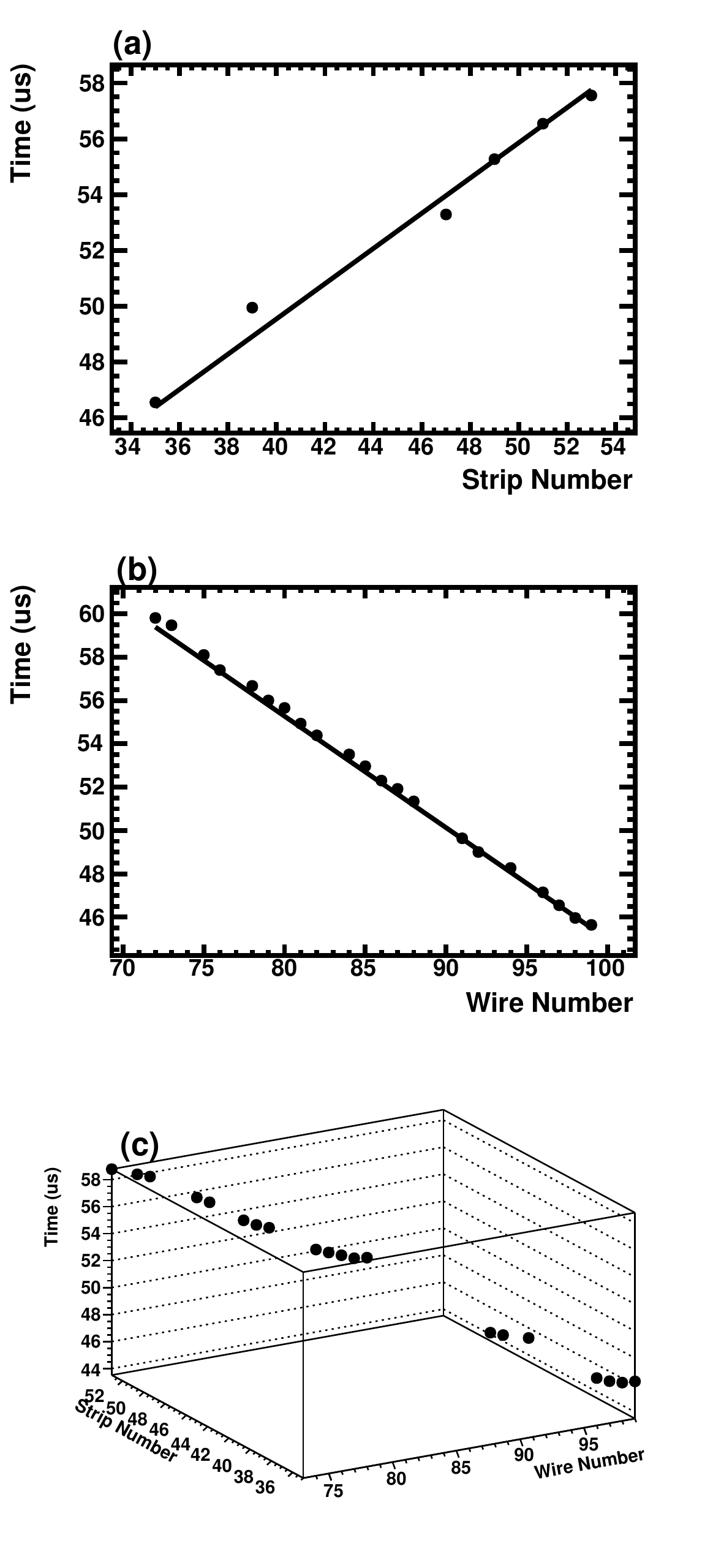}
\caption{An example event, comprising hits found in the (a) strip readout plane, the (b) wire readout plane, and (c) three dimensional hits deduced from those.} \label{fig:tracks}
\end{figure}

After baseline subtraction, a software threshold is applied to each waveform to search for ``hits'' due to ionization tracks. Since the data set predominantly contains single elastic scatter events, we assume in this analysis that each waveform contains only one hit. The hit time is found by performing a quadratic fit over a $20$ sample range centered upon the maximum amplitude sample in the waveform. A set of hits thus found for the wire and strip planes in a typical event are shown in Figures~\ref{fig:tracks}(a)\&(b).

The procedure described above finds the projection of the 3D ionization track onto the pair of crossed two dimensional (2D) readout planes. We are, of course, interested in reconstructing the full 3D track. The waveform timing information is used to form coincidences between strip and wire plane hits. The waveforms from each wire and strip containing a valid 2D hit are multiplied together, and a hit search similar to that described above is performed, yielding a set of 3D hits (Fig.~\ref{fig:tracks}(c)). Waveforms that overlap in time produce combined traces with clear ``hits''; those that do not overlap in time produce flat combined traces with no overlapping ``hit''. 


 
\subsection{Tracking}

To determine track parameters, we begin with the assumption that proton recoil tracks are linear, i.e. will exhibit little straggling, which is appropriate given the light mass of the stopping medium. To identify linear features we apply a discrete Hough transform \cite{Hough1,Hough2}, since this technique is robust against noise. If a linear feature is found, hits that lie further than $\approx1$~cm from it are discarded. 

We implement the Hough transform with fairly coarse bins for computational efficiency and operational simplicity. To realize the full tracking resolution of the TPC we then apply a three dimensional least squares fit to the remaining hit coordinates to extract the final (continuous) track parameters. These are expressed as two angles; an elevation angle referenced from the horizontal plane that runs through the center of the TPC ($\pm 90^{\circ}$, denoted ``Elevation''), and an azimuth angle referenced from the axis of the TPC vessel perpendicular to the charge readout plane ($\pm 180^{\circ}$, denoted ``Phi''). Examination of the track ionization profile determines the particle orientation along the track.


\subsection{Ionization Measurements}
  
\begin{figure}[tb]
\centering
\includegraphics*[width=2.2in,angle=90]{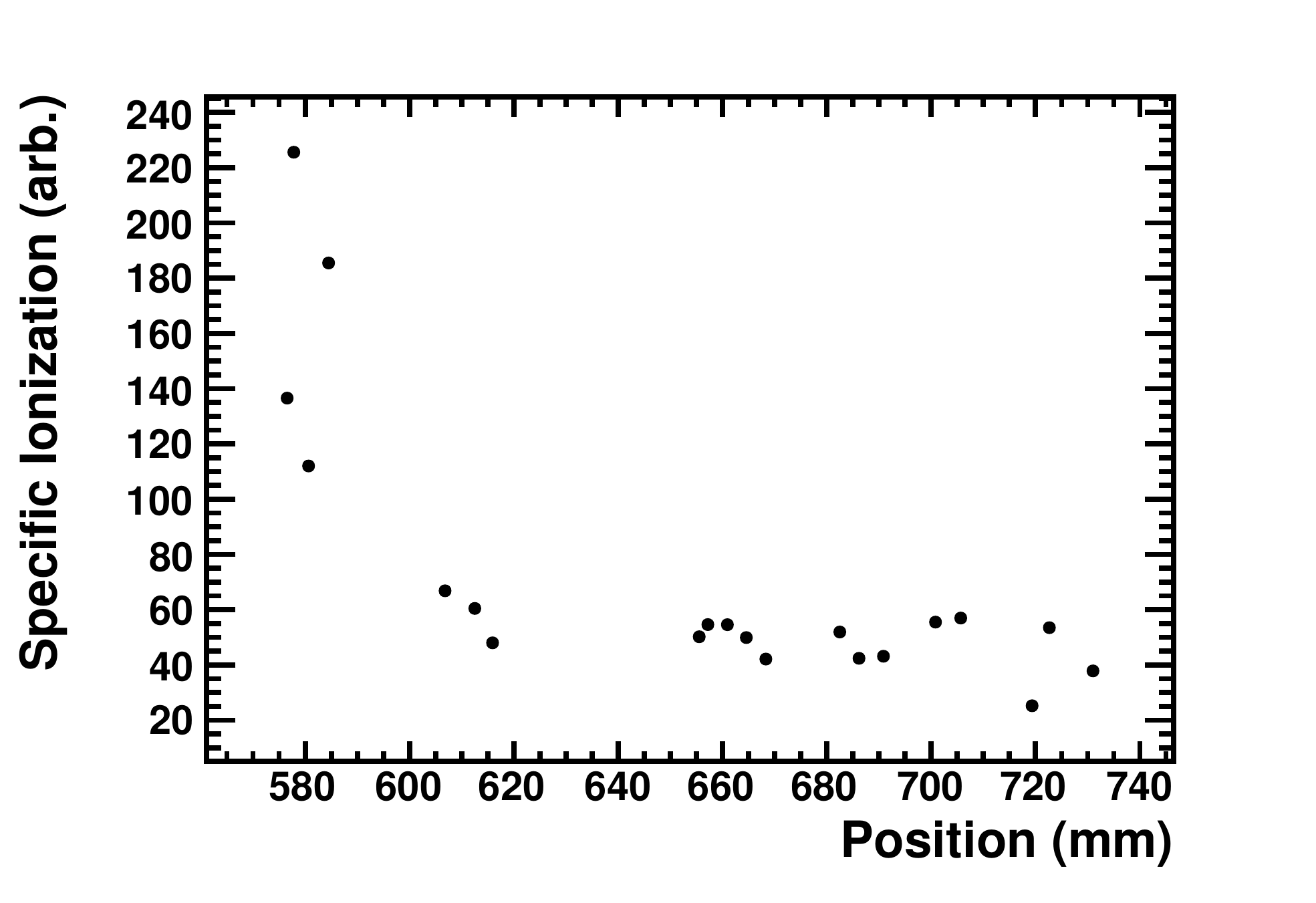}
\caption{The ionization profile of the event shown in Fig.~\ref{fig:tracks}.} 
\label{fig:ionization}
\end{figure}

With knowledge of the track parameters, the linear position of each hit along the track can be easily calculated. The wire charge measurements are appropriately scaled to yield the specific ionization along the particle track (Fig.~\ref{fig:ionization}). Examination of the track and identification of the position of the Bragg peak allows breaking of the two-fold pointing degeneracy. The point of greatest specific ionization is designated the Bragg peak, and the remaining ionization values are averaged to yield a quantity that we denote "Mean Tail Ionization." These two quantities are used to assess track quality and to perform particle identification. Integration of the specific ionization along the length of the profile is used to determine the particle energy. 

\subsection{Uniform Efficiency and Track Quality Cuts}
\label{sec:quality}

We examine several quantities to assess the quality of the tracks that are reconstructed, and to ensure a uniform angular acceptance. First, we exclude track angles and track lengths at which the data acquisition system is not uniformly efficient as a function of direction. In practice, this means excluding track angles within $10^{\circ}$ to the drift axis of the TPC ($80<|$Phi$|<100$), track angles at high elevations ($|$Elevation$|>80$), and requiring that particle tracks be at least $20$~mm long. The former requirement reduces the field of view by $\approx 10\%$, while the later results in a loss of efficiency of about $25\%$ when measuring fission neutron sources. We again note that this field of view reduction is a feature of this particular prototype implementation, rather than of TPCs in general. 

\begin{figure}[tb]
\centering
\includegraphics*[width=2.2in,angle=90]{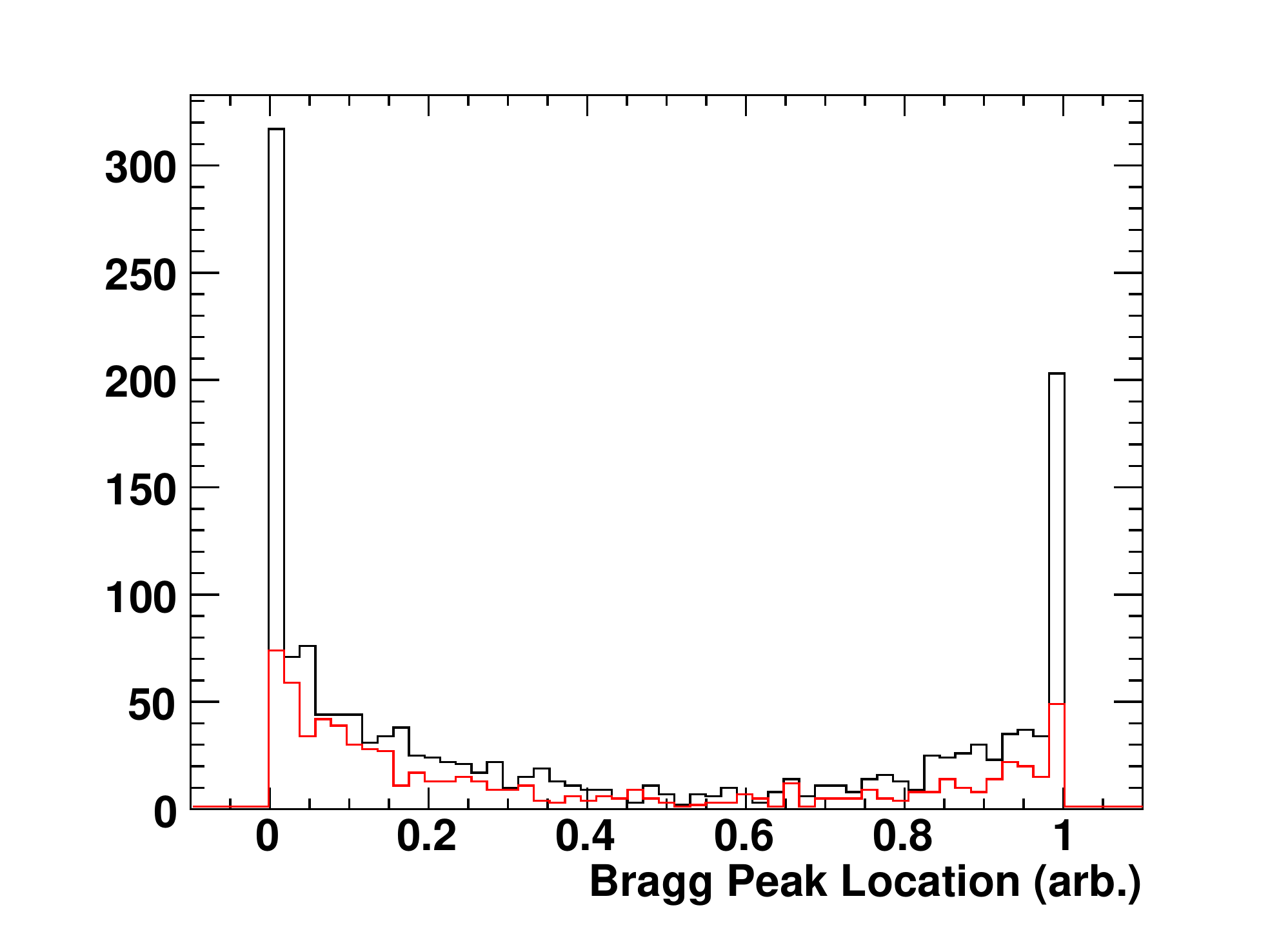}
\caption{The distribution of the Bragg Peak Location (BPL) parameter for a background run (red) and a run with a neutron source (black). Many background events do not have a Bragg peak at their terminus (BPL $\approx 0$~or~$1$) and can therefore be rejected.} 
\label{fig:bpl}
\end{figure}

We then examine the ionization profile of events to determine whether they follow the expected form of stopping charged particles, i.e. whether the event ionization profile resembles a Bragg curve. The location of the Bragg peak is determined, and the event rejected if it does not fall within the first or last 10\% of the track length (Fig.~\ref{fig:bpl}). 

\subsection{Particle Identification}

\begin{figure}[!tb]
\centering
\includegraphics*[width=3in]{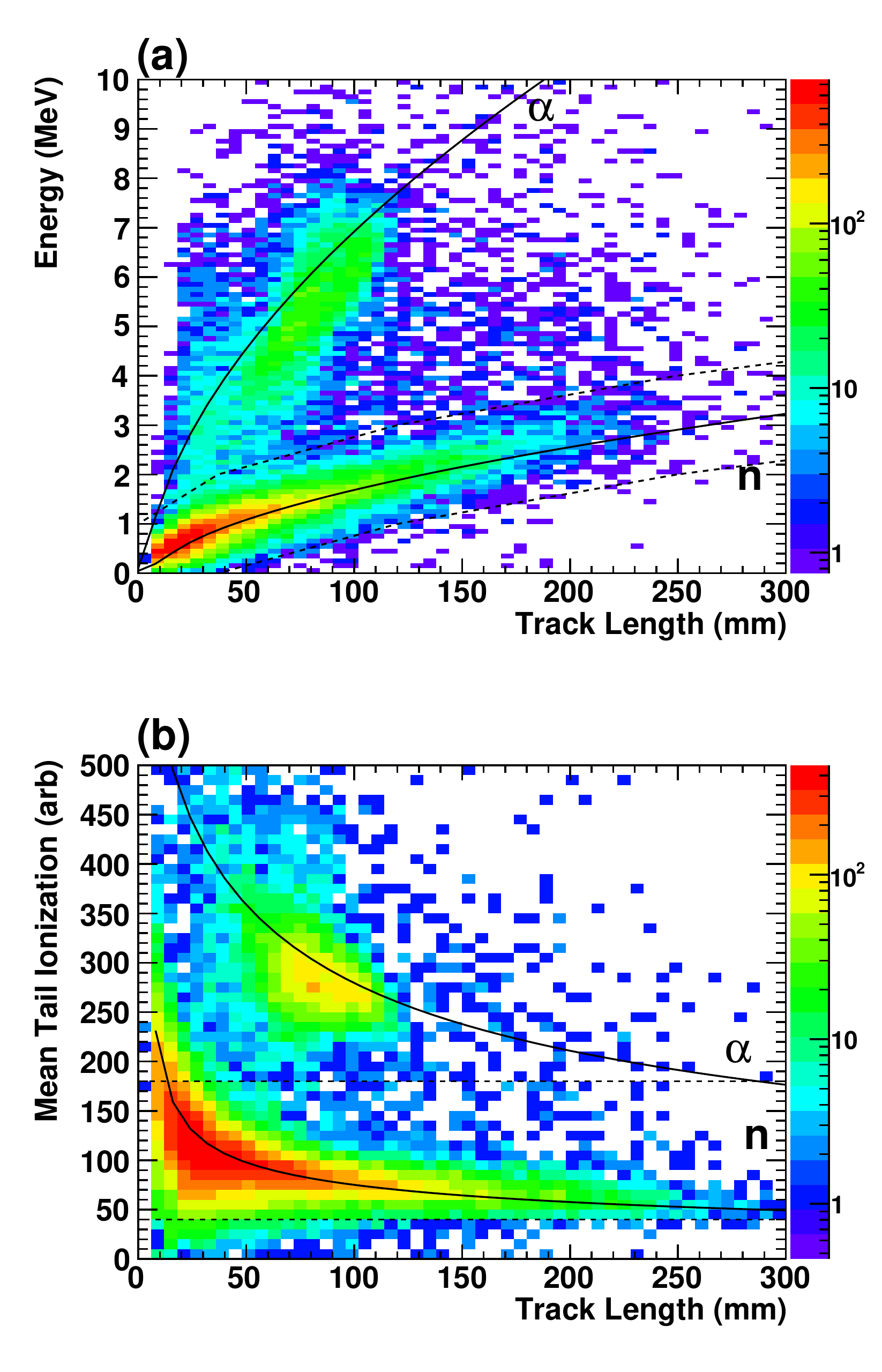}
\caption{The event track length is compared to (a) the particle energy and (b) Mean Tail Ionization, in order to distinguish particle type. In these plots data from separate neutron and alpha source calibrations runs are combined. The solid lines are predictions calculated using the SRIM package, while the dashed lines indicate the proton selection regions.} 
\label{fig:tle}
\end{figure}

Information from the ionization profile also allows identification of tracks created by protons, and rejection of gamma and alpha particle tracks (due to the inadvertent use of a considerable area of lead solder on the interior of the field cage, there is a few Hz alpha particle rate in TPC from $^{210}$Pb decay). The first step of our particle identification technique is examination of Energy divided by Track Length (Average Specific Ionization). However, when short tracks dominated by the Bragg peak are considered, this quantity is a poor discriminant. This is demonstrated in Fig.~\ref{fig:tle}(a), where data from neutron and alpha source calibration runs are combined, and compared to a prediction calculated using the SRIM package~\cite{SRIM}. At short length, the alpha and proton bands converge. 

To provide robust particle discrimination at short track length, we also compare the Mean Tail Ionization quantity to the track length (Fig.~\ref{fig:tle}(b)). For long tracks, this quantity is approximately independent of length. For shorter tracks, dominated by the rise in specific ionization approaching the Bragg peak, the Mean Tail Ionization begins to rise, but distinct proton and alpha bands remain.

\begin{figure}[tb]
\centering
\includegraphics*[width=2.2in,angle=90]{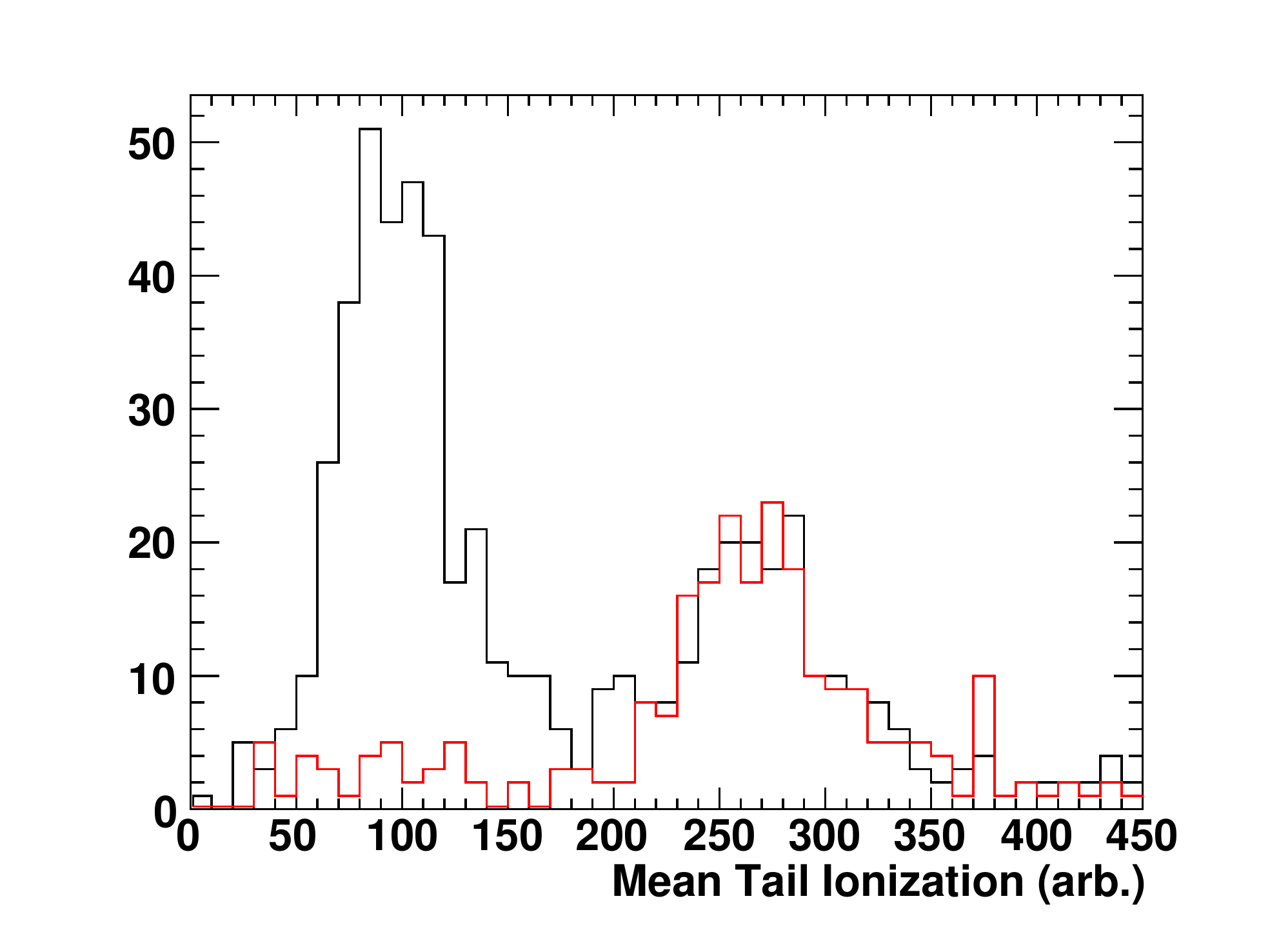}
\caption{The mean ionization found in the ``tail'' of tracks is used to identify particle type. Here we compare a run with a neutron source $8.7$~m from the TPC (black) to a background run of equal duration (red). The peak at $\approx 100$ is due to recoil protons while that at $\approx 250$ is due to alphas.} 
\label{fig:tm}
\end{figure}

A simple cut on the Mean Tail Ionization variable ($40<$ Mean Tail Ionization $<180$)  provides good discrimination (dashed lines in Fig.~\ref{fig:tle}(b)), albeit at the expense of some efficiency at short lengths.  To provide a sense of average signal and background rates encountered in this device, Fig.~\ref{fig:tm} compares the Mean Tail Ionization distribution of a background run and a run with a neutron source placed $8.7$~m from the TPC.

\section{TPC Calibration}
\label{sec:calibration}

A variety of measurements were made to characterize and calibrate the TPC in preparation for neutron measurements.

\subsection{Drift Speed Measurement}

Reconstruction of tracks in three dimensions requires measurement of the the speed at which ionization charge drifts to the readout plane. Our prototype TPC did not include a laser calibration system or external lifetime monitor; therefore the following calibration method was used.  A neutron source was placed approximately 10 meters from the chamber at $10^{\circ}$ increments in the horizontal plane.  The drift speed loaded into the reconstruction software was then varied through reasonable values until the sources were measured to be at the correct locations.  The drift speed depends on the pressure, temperature and gas composition but was typically around 1cm/$\mu$s for the data presented here.  The Magboltz program~\cite{Magboltz} was used to calculate the drift velocity via numerical integration of the Boltzmann transport equation, yielding qualitative agreement with the experimental measurement.

\subsection{Channel-by-Channel Gain Correction}

The readout channel gains were calibrated relative to one another using neutron data. Event ionization profiles, like that shown in Fig.~\ref{fig:ionization}, were fit to an analytic approximation of the form of the Bragg curve \cite{bragg},
\begin{equation}
\frac{dE}{dx}(x) \approx \frac{a}{(R-x)^{1-\frac{1}{p}}}\label{equ:bragg},
\end{equation}
where $R$ is the range of the particle, $a$ is a constant, and $p=1.8$ for protons. The scale factor that would have to be applied to each point on the track to match the fit was calculated, and averaged over many events to calculate a gain correction factor. Typical corrections were of magnitude $10-20\%$. This procedure was repeated iteratively several times, with the correction factor for one reference channel being held fixed. This calibration was checked by applying a pulsed electrical signal directly to the Frisch grid. This yielded similar channel-by-channel corrections, implying that the major source of gain variation was variation among the preamplifiers. 

\subsection{Ionization Loss due to Electronegative Contaminants}

Outgassing from detectors components can introduce electronegative impurities that absorb primary ionization electrons as they drift to the charge readout plane. The effect of outgassing was measured by sealing the gas vessel and following the system gain over a period of $5$~days. A gain degradation of almost $7\%$/day was observed. When a constant gas flow is maintained, an equilibrium is reached at which impurities are being added by outgassing at the same rate at which the gas flow is removing them. With a flow of $60$~liters/hour it was found that this equilibrium was reached only $12$~hours after introducing the counting gas to an evacuated TPC vessel, with a net reduction in gain of only $4\%$.  Therefore, at least $12$~hours was allowed to elapse before any data taking. We note that no effort was made to select materials that exhibit low outgassing for this laboratory prototype. This suggests that use of materials that exhibit low outgassing, e.g. metals, kapton, and non-porous ceramics, could improve this situation considerably, and possibly allow for sealed operation.

\subsubsection{Correction for Gain Variation due to Temperature Variation}

The small effect of ambient temperature variations on gain was also measured. Given that the TPC is operated in a constant pressure mode, temperature variations result in gas density variations, which in turn result in gain variations \cite{Blum_n_Rolandi}:
\begin{equation}
\frac{dG}{G} \propto \frac{d\rho}{\rho} \propto -  \frac{dT}{T}.
\label{equ:gain_density}
\end{equation}
Several resistive temperature devices mounted upon the strip plane were used to correlate changes in gas temperature with changes in gas gain over the course of several days. A linear relationship between temperature and gain was observed, with a coefficient of $1.5\%/^{\circ}$C. Given that typical diurnal temperature variations in our laboratory are $\approx 0.2^{\circ}$C, this is a small effect, but nonetheless a temperature derived gain correction is applied. The application of thermal insulation to the exterior of the gas vessel would presumably reduce this small effect yet further.

\subsection{Energy calibration}

\begin{figure}[tb]
\centering
\includegraphics*[width=2.2in,angle=90]{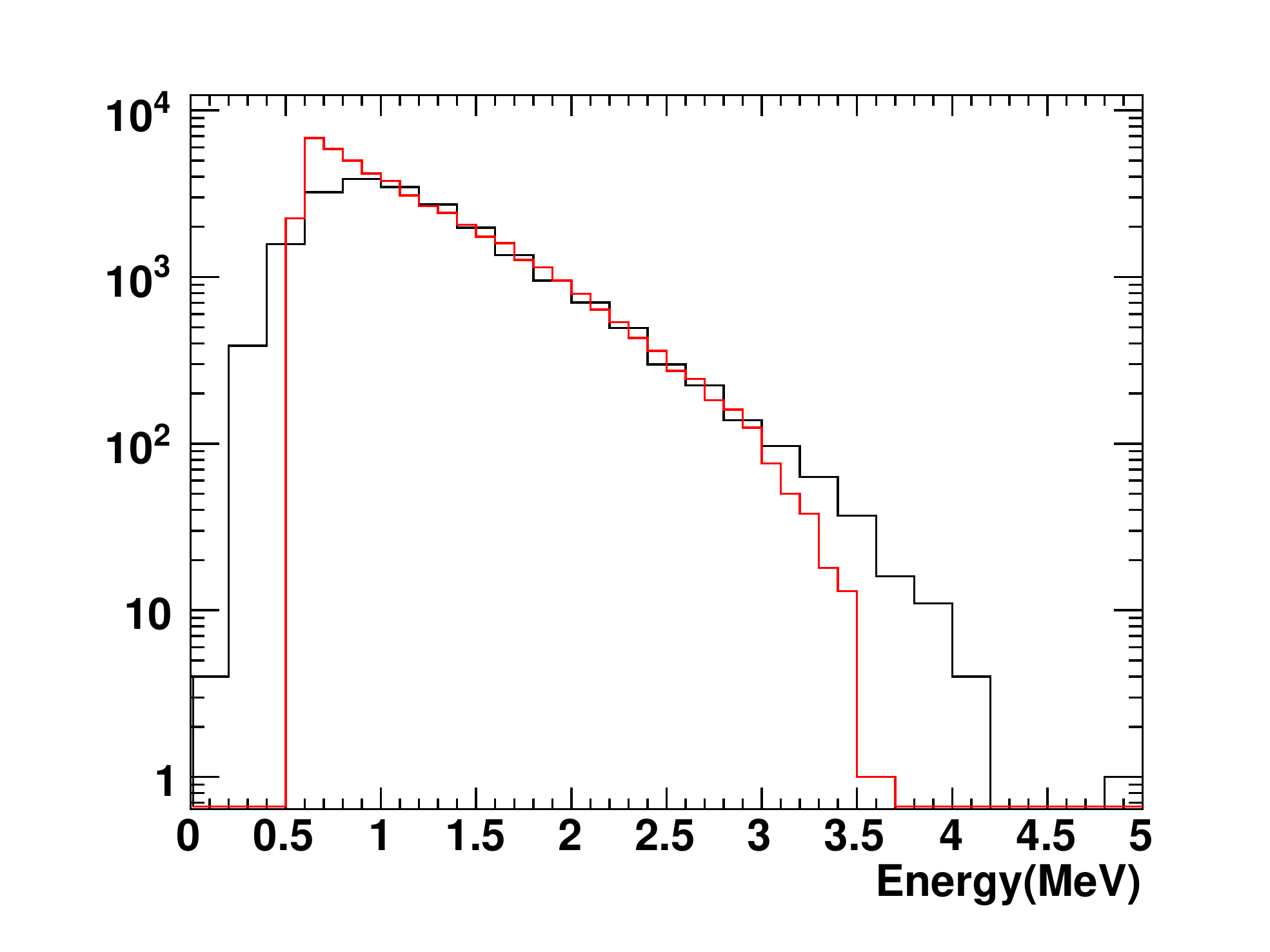}
\caption{The measured proton recoil spectrum due to a fission neutron source (black) compared to a Monte Carlo prediction (red).} 
\label{fig:R_S}
\end{figure}

The conversion factor between the total measured ionization and energy for proton recoil events was determined using measured track lengths. The relationship between proton track length and proton energy was deduced for the gas composition and pressure used here using SRIM~ \cite{SRIM}. The length/energy relationship thus determined was parametrized using a 2nd order polynomial, which was in turn fit to the proton data presented in Fig.~\ref{fig:tle}(a) to yield the energy scale calibration in terms of MeV. The recoil proton spectra due to a $^{252}$Cf fission neutron source can be seen in Fig.~\ref{fig:R_S}. This is compared to a simple Monte Carlo prediction that includes only those recoils that are fully contained within the active TPC volume. This condition effectively applies a high energy cutoff. 

\section{TPC Design and Performance Simulation}
\label{sec:simulation}

In order to assess the characteristics and pointing ability of our detector design, a Monte Carlo simulation of the TPC has been carried out using the MCNP \cite{MCNP} and MCNP-Polimi codes \cite{Polimi}. The Monte Carlo model incorporates the main features of the detector design (stainless steel chamber, field cage, and ionization gas), but not, in this implementation, laboratory walls or the intervening atmosphere. Simulation predictions of the angular resolution that can be achieved with the single elastic scatter technique are compared directly to data in Sec.~\ref{sec:direction}.


We also used this simulation to predict the magnitude of scattering from the gas vessel and field cage. It is predicted that $\sim92\%$ of neutrons that scatter in the TPC active volume experience their first scatter in that volume, with the remainder predominantly scattering first off the steel of the vessel. The small fraction of neutrons whose path is altered before interaction in the active volume is entirely acceptable in this proof of principle demonstration, and could be reduced by careful tailoring of the gas vessel design to the exact parameters required by a particular application.

\section{Neutron Measurements}

\subsection{Directional Response}
\label{sec:direction}

\begin{figure*}[htb]
\centering
\includegraphics*[width=6in]{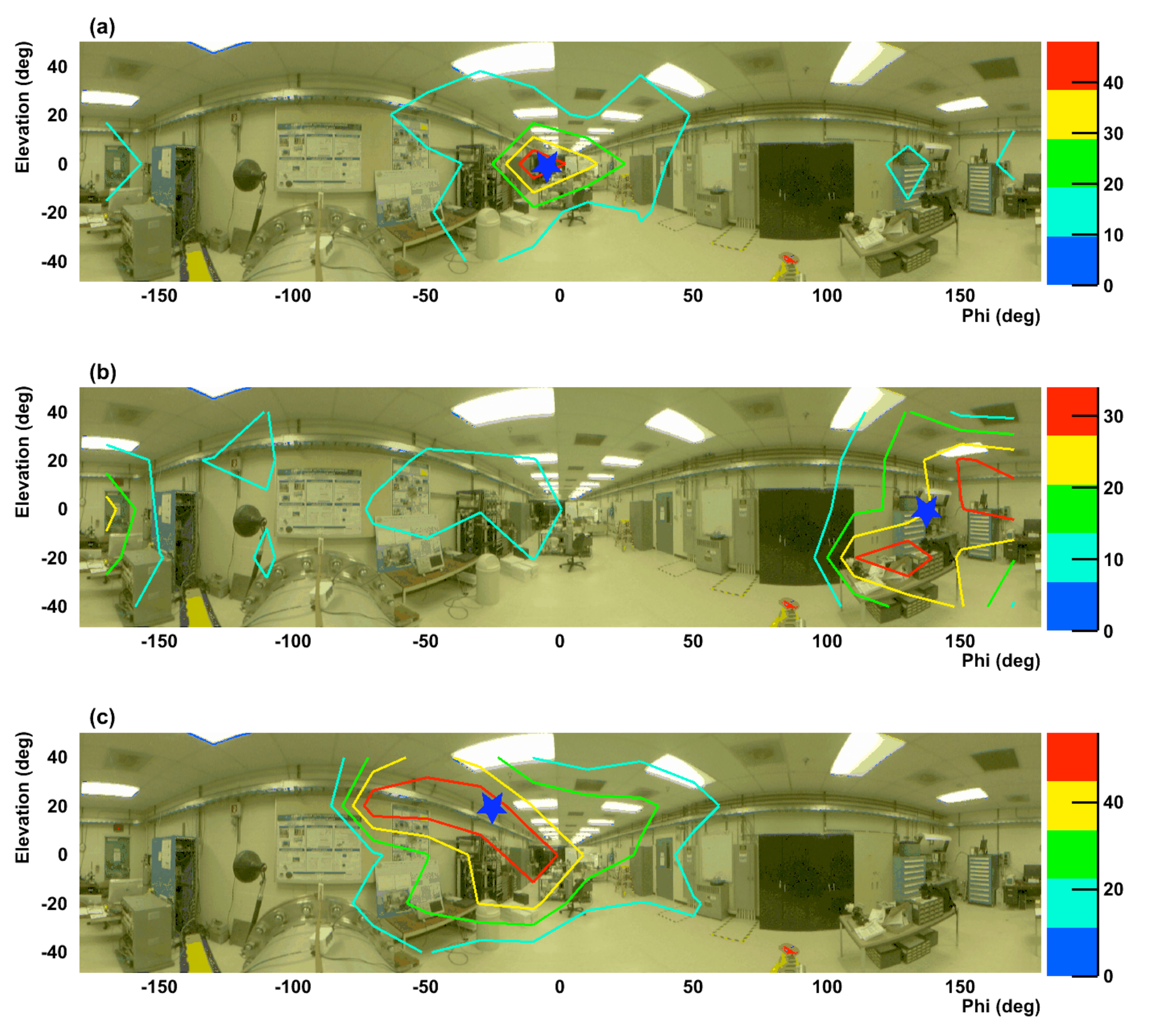}
\caption{Contour plots demonstrating the pointing ability and wide angular field-of-view of the TPC with a $^{252}$Cf source placed at various positions within the experimental laboratory. The blue star denotes the source position. The source standoffs were (a) $17.7$~m, (b) $5.7$~m, and (c) $4.0$~m. Note that the TPC is senstive beyond Elevation~$=\pm40$; this representation of the data is limited by the abilities of the panoramic camera used.} 
\label{fig:panorama}
\end{figure*}

The imaging performance of the TPC was assessed by placing a $60~\mu$Ci $^{252}$Cf fission neutron source at a variety of angular positions and standoffs (as large as $17.7$~m). In Fig.~\ref{fig:panorama} we show the response of the TPC to three such experiments, demonstrating the broad field of view of this device. As discussed in \cite{nTPC2}, there is a direct correlation between the peak of angular distribution of recoil protons and the direction of incident neutrons. Thus, we bin the reconstructed direction of detected recoil protons, and overlap a contour plot of the resulting distribution upon a panoramic photograph of the experimental laboratory.  The correspondence of that distribution with the source position can clearly be observed.

\begin{figure}[tb]
\centering
\includegraphics*[width=3in]{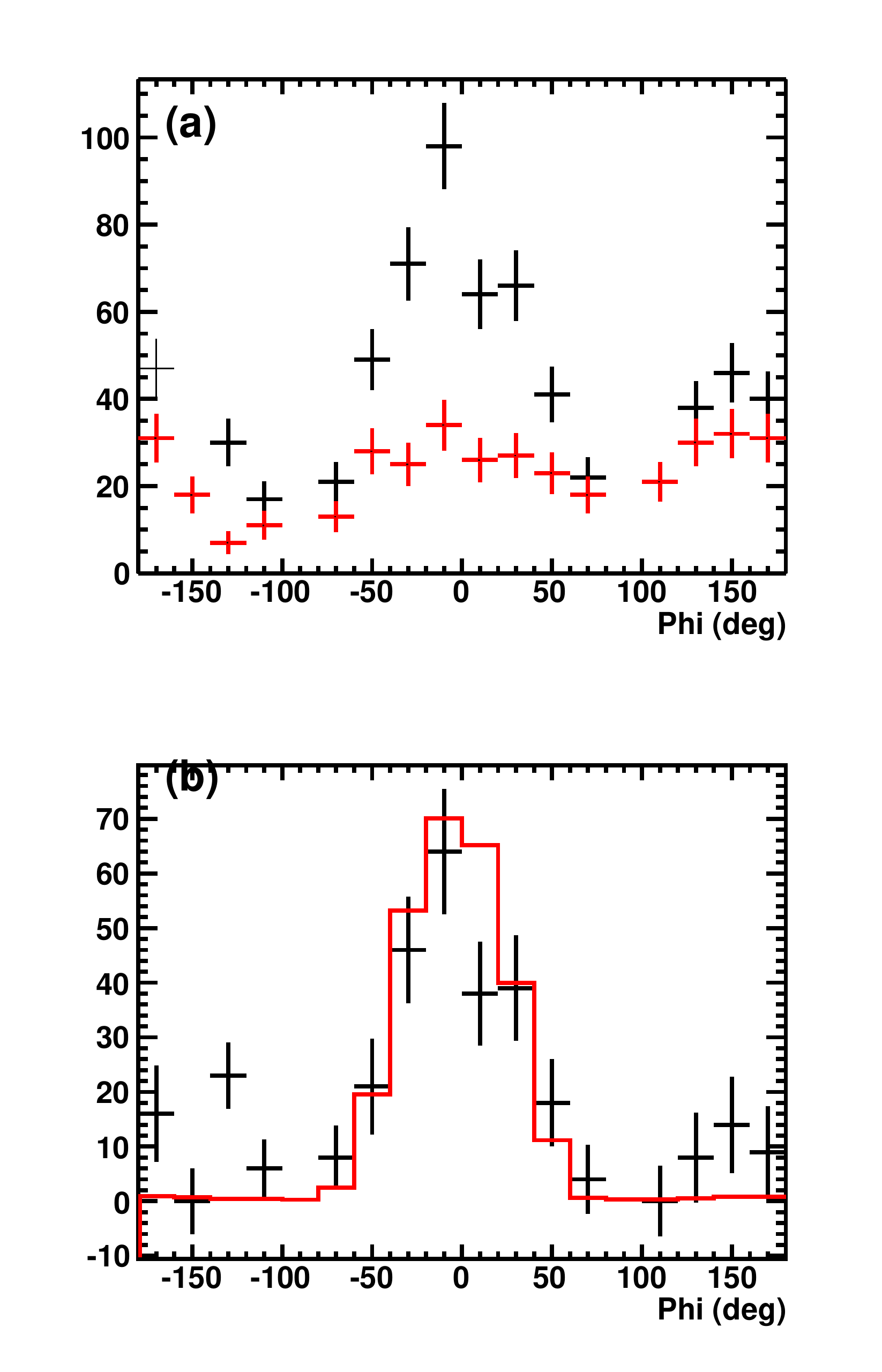}
\caption{(a) The recoil azimuth angle distribution of events acquired with a neutron source $17.7$~m from the TPC (black), and background (red). (b) Background subtracted azimuth angle distribution of events due to neutrons from the $17.7$~m distant source, compared to an MCNP prediction (red).} 
\label{fig:distant}
\end{figure}

To demonstrate the benefit of directional detection in increasing signal to background, consider the data presented in Fig.~\ref{fig:distant}(a), comparing data taken with the neutron source $17.7$~m from the TPC and background. In this example, the excess of events due to the source is less than the background, when integrated over the entire angular range. However, the directional detection technique demonstrated here yields a considerably greater signal/background ratio when that integral is limited to a narrower angular range centered upon the source direction. 

\begin{figure}[!tb]
\centering
\includegraphics*[width=2.2in,angle=90]{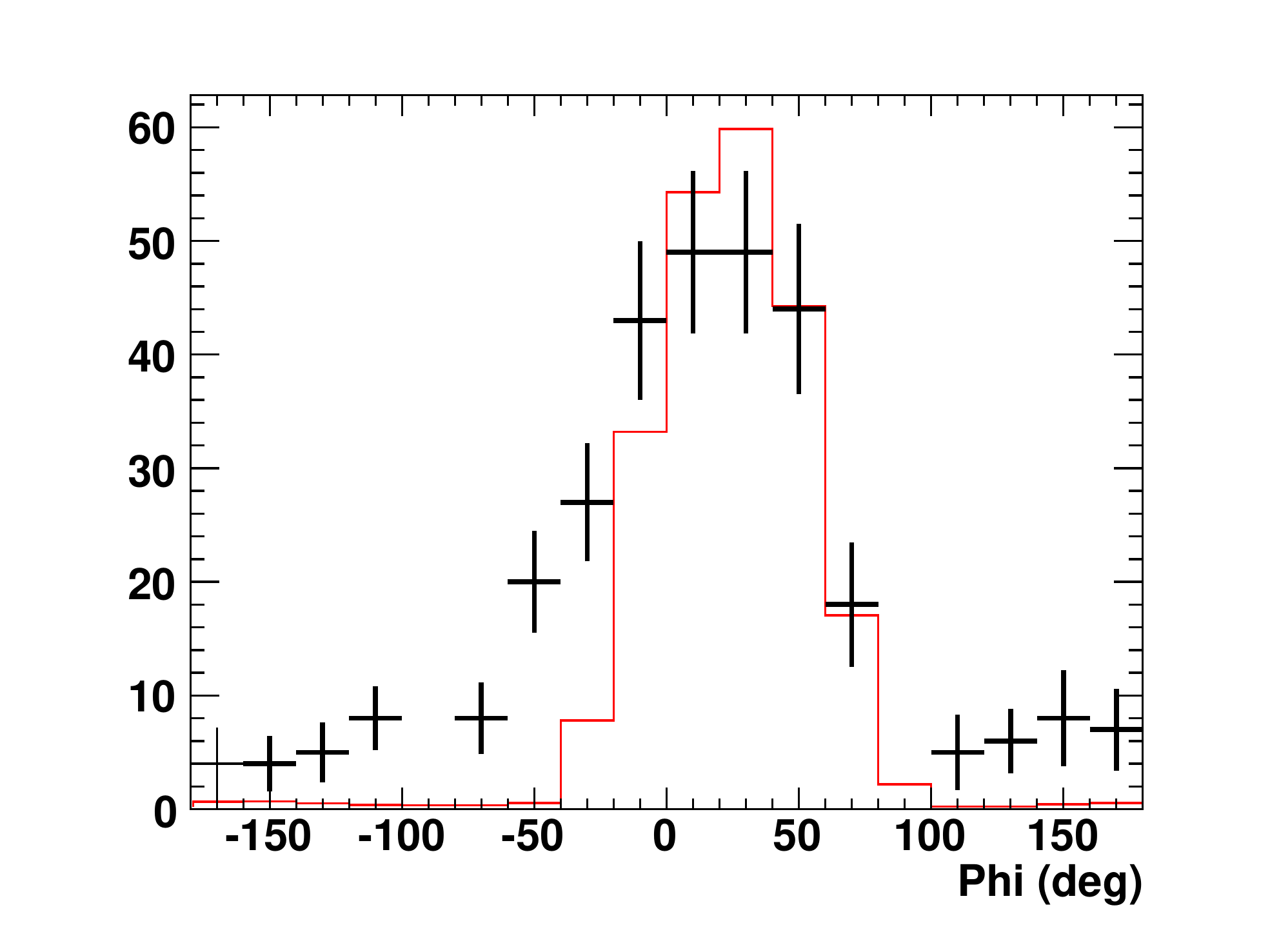}
\caption{Background subtracted azimuth angle distribution of events due to neutrons from a neutron source placed close to laboratory walls, compared to an MCNP prediction (red)} 
\label{fig:data_MCNP}
\end{figure}

The angular resolution, i.e. ability to resolve closely spaced sources, of the simple source direction reconstruction method presented here (mean position of the recoil distribution) is just the width of that distribution. We draw the readers attention to the apparent difference in the width of the recoil distribution between the various experiments shown in Fig.~\ref{fig:panorama}. We tentatively attribute this to the effects of room scatter -- in Figs.~\ref{fig:panorama}(b)\&(c) the neutron source was close to concrete walls, cabinets, etc, while it was far from walls as was possible in Fig.~\ref{fig:panorama}(a). We therefore expect the data presented in Fig.~\ref{fig:panorama}(a) to be a better representation of the inherent angular resolution of the TPC, although further experimentation in a large, empty experimental location would be required to fully demonstrate this. This interpretation of the differences in recoil width is supported by comparisons of the measured widths to that determined from the MCNP simulation. This comparison is made in Fig.~\ref{fig:distant}(b) for an experiment distant from walls, while that displayed in Fig.~\ref{fig:data_MCNP} was obtained with the neutron source adjacent a laboratory wall. 
 
\begin{figure}[tb]
\centering
\includegraphics*[width=3in]{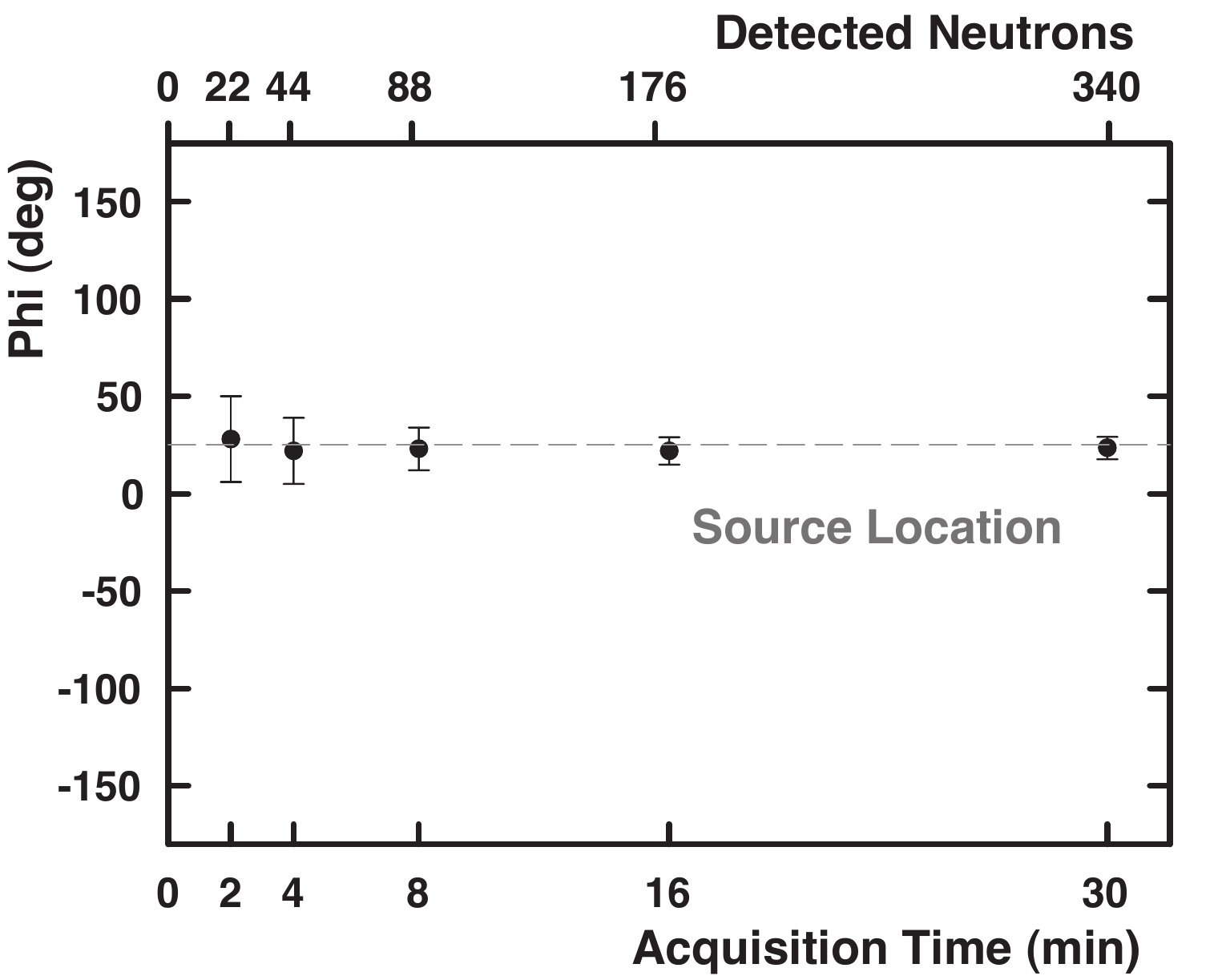}
\caption{The precision with which the direction of a neutron source located $8.7$~m from the TPC can be determined improves as increased statistics are accumulated. The error bars correspond to the standard deviation of the azimuthal angle of the recoil protons for a source placed in a horizontal plane of the detector.} 
\label{fig:Pointing}
\end{figure}

While the recoil proton distribution itself is fairly broad, this is not an indication of the accuracy with which the mean direction of that distribution, i.e. the neutron source direction, can be determined. This is demonstrated in Fig.~\ref{fig:Pointing}, where the direction within the TPC horizontal plane of a $^{252}$Cf source is determined as a function of dwell time via a Gaussian fit to the recoil distribution. As neutron counts are accumulated, the estimate of the \textit{mean position} of the recoil distribution improves considerably. Detection of even a few tens of neutrons results in a significant reduction of the search space, compared to a non-imaging detector. 

\begin{table*}[htb]
\caption{Estimates of the TPC neutron detection efficiency. Errors quoted reflect only counting statistics and neutron source strength uncertainty. 
The neutron flux reaching the TPC active area from the source at 8.7m distance is $32 \pm 3$~Hz and  $8 \pm 0.8$ at $17.7$~m, assuming no scattering or attenuation in the intervening material or atmosphere.} \label{tab:eff}
\begin{tabular}{l l c c c c} \hline
Distance&Analysis Cut&Neutron + Bkgd Rate (Hz)&Bkgd Rate (Hz)&Neutron Rate (Hz)&Efficiency (\%)\\ \hline 
$8.7$~m&Raw& $0.85\pm0.02$& $0.40\pm0.02$& $0.44\pm0.03$&$1.4\pm0.2$    \\
&Particle ID& $0.41\pm0.02$& $0.055\pm0.006$& $0.36\pm0.02$&$1.1\pm0.1 $   \\ 
&Track Quality& $0.24\pm0.01$& $0.024\pm0.004$& $0.22\pm0.01$&$0.69\pm0.08$     \\
&Uniform Efficiency& $0.19\pm0.01$& $0.019\pm0.003$& $0.17\pm0.01$&$0.55\pm0.07$     \\\hline
$17.7$~m&Raw& $0.377\pm0.004$& $0.350\pm0.004$& $0.027\pm0.005$&$0.35\pm0.08$    \\
&Particle ID& $0.064\pm0.002$& $0.040\pm0.001$& $0.024\pm0.002$&$0.32\pm0.04 $   \\ 
&Track Quality& $0.033\pm0.001$& $0.019\pm0.001$& $0.015\pm0.001$&$0.19\pm0.03$     \\
&Uniform Efficiency& $0.027\pm0.001$& $0.015\pm0.001$& $0.012\pm0.001$&$0.16\pm0.02$     \\\hline
\end{tabular}
\end{table*}

\subsection{Neutron Detection Efficiency}

In Table~\ref{tab:eff}, we examine event rates and the effect of various analysis steps in order to estimate the neutron detection efficiency of the TPC. Data collected with a $^{252}$Cf fission neutron source placed $8.7$~m and $17.7$~m from the TPC vessel are compared to background runs of equal length in each case. We use these data sets, since they are respectively likely to either overestimate (due to room scatter) or underestimate (due to several inches of unquantified intervening material) the efficiency. We estimate the number of neutrons produced by the source to be $(3 \pm 0.3) \times 10^5/$s. The large uncertainty results from the advanced age of the source ($24$~years), and the consequent uncertainty as to the contribution of $^{250}$Cf spontaneous fission to the neutron emission rate~\cite{old_cf}. Again, further experimentation in a large, empty experimental location will be required to gain a more accurate estimate.

With no analysis cuts applied, an excess of events due to the neutron source equivalent to a $\approx 1\%$ neutron interaction efficiency is observed. Application of all analysis cuts results in a $\approx 50\%$ reduction in efficiency. These experimental results are in reasonable agreement with calculations of the expected efficiency obtianed via both a first principle calculation and the MCNP simulation. These yield predictions for the raw neutron interaction efficiency of $\approx 1.1\%$ and for the ultimate efficiency after selection cuts of $\approx 0.4\%$. 

We also note the low measured neutron background rate of $0.015\pm0.001$~Hz, which implies a background flux of $\approx 8\times10^{-4}$ cm$^{-2}$s$^{-1}$, assuming the predicted detection efficiency of $0.4\%$. This is in reasonable agreement with quoted values for the integrated fast neutron background rate ($\approx1.5\times10^{-3}$ cm$^{-2}$s$^{-1}$ \cite{Gordon}) for fission neutron energy range ($0.5 - 10$~MeV), especially given the large uncertainties related to the neutron shielding provided by and cosmogenic production within the concrete walls of the laboratory. 

\subsection{Non-neutron Background Sensitivity}

\begin{figure}[tb]
\centering
\includegraphics*[width=2.5in,angle=90]{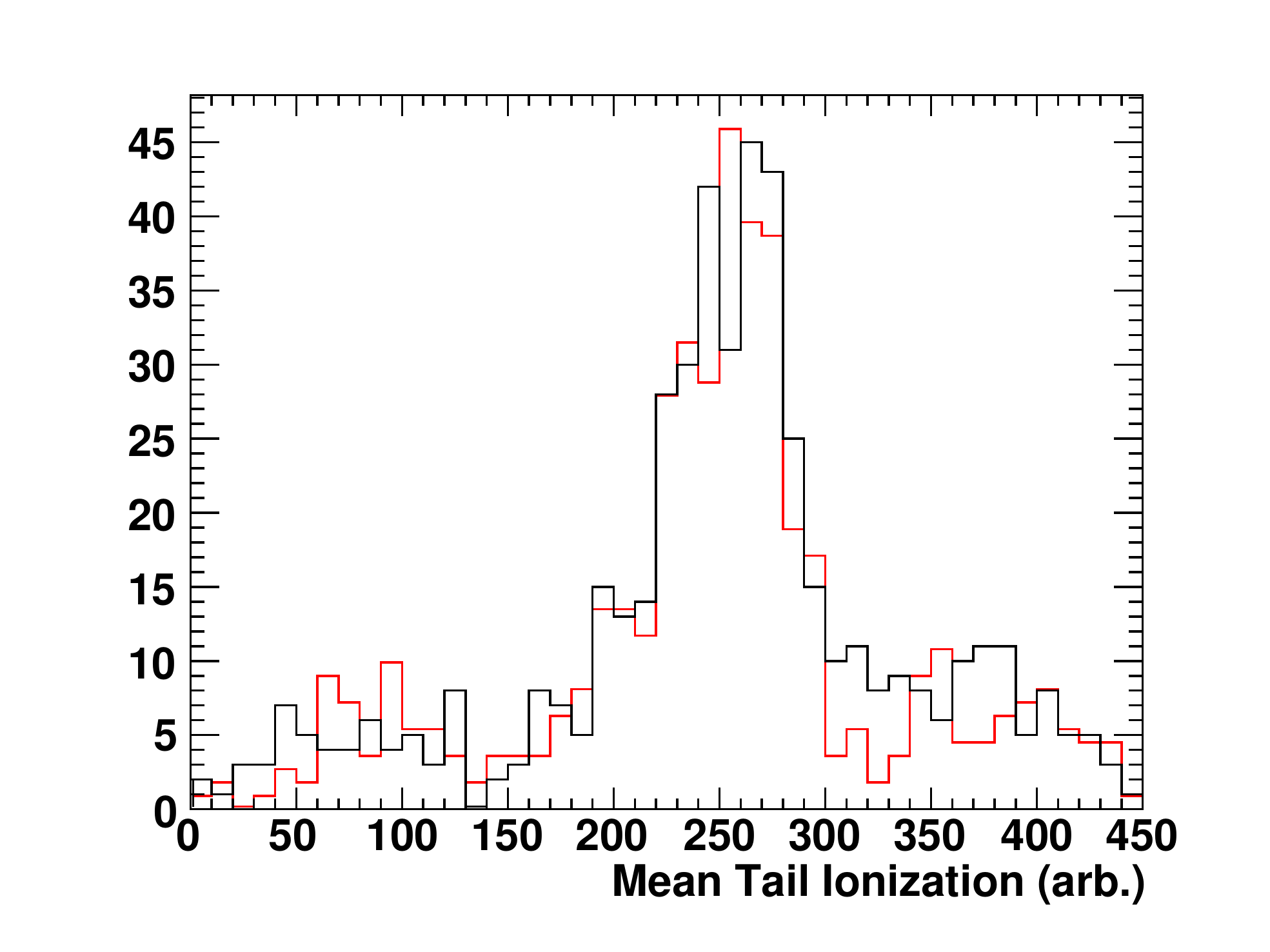}
\caption{The mean ionization measured in the track tail is compared for a run with a $2$~MBq $^{60}$Co source beside the TPC (black) and a background run of equal duration (red). No appreciable gamma ray response is observed. The peak at $\approx250$ is due to alpha particle tracks within the TPC, while if they were measured compton electrons due to gamma rays would be expected to be less than 10.} 
\label{fig:gamma}
\end{figure}

To demonstrate the inherent gamma ray insensitivity of this device, we placed a $2$~MBq $^{60}$Co source beside the TPC. It was estimated that the interaction rate of $^{60}$Co gamma rays within the TPC gas was $\approx 400$~Hz. As can be seen in Fig.~\ref{fig:gamma}, no appreciable gamma response was measured. In the neutron selection region the event rate without the gamma source present was $0.022 \pm 0.002$~Hz, agreeing with that measured with the source present ($0.020 \pm 0.002$~Hz). We therefore estimate the gamma rejection of this device to be $> 10^6$.

We have also examined the response of the TPC to microphonics, by striking the TPC vessel with a metallic object with enough force to set the readout wires in motion. Spurious events were generated, most likely by wire oscillation, but these did not pass the neutron selection cuts. The net result was a reduction is livetime, due to a saturation of the relatively low throughput DAQ system, e.g. tapping on vessel $6$~times per minute resulted in a $25\%$ livetime reduction. A higher throughput DAQ and a more rigid charge readout system, e.g. a LEM or thick GEM~\cite{LEM}, would improve this considerably. Such effects are not observed during normal laboratory operation, e.g. personnel moving around the device.

\section{Conclusion}
\label{sec:conclusion}

In conclusion, a prototype directional detector for fast neutrons based on a H$_2$ filled TPC has been designed, constructed, and characterized. The detector exhibits favorable characteristics, such as the field of view nearly equal to the entire $4\pi$ solid angle, low sensitivity to gamma-ray background, and the ability to rapidly localize neutron source directions. The ability of the TPC to point to a fission neutron source has been experimentally demonstrated in a laboratory environment with the efficiency of approximately $0.5\%$, and at several tens of meter standoff. 

Future efforts will focus on the exploration of higher resolution imaging capabilities using multiple neutron scatters and the incorporation of energy information into the reconstruction for single scatters, and on the optimization of detector design for field applications related to the detection of special nuclear material. In particular, lightweight vessels operated at atmospheric pressure, using for example isobutane, seem attractive for mobile search applications. 

\section*{Acknowledgements}

We thank M. Howe for his assistance with the ORCA acquisition software. We are grateful to L. Rosenberg and A. Bernstein for early contributions to this project.

This work was supported by DOE/NA-22.

LLNL-JRNL-428129.

This work was performed under the auspices of the U.S. Department of Energy by Lawrence Livermore National Laboratory in part under Contract W-7405-Eng-48 and in part under Contract DE-AC52-07NA27344.

\end{document}